\documentclass[twocolumn,showpacs,preprintnumbers,amsmath,amssymb]{revtex4}

\usepackage{graphicx}
\usepackage{dcolumn}
\usepackage{bbm}

\newcommand{ \mo }[1]{ \hat{ #1 } }
\newcommand{ \bra }[1]{ \langle #1 | }
\newcommand{ \ket }[1]{ | #1 \rangle }

\newcommand{ \mv }[1]{ \mathbf{ #1 } }

\newcommand{ \abs }[1]{ \left| #1 \right| }

\newcommand{ \sE }{ \mathbb{E} }

\newcommand{ \Tr }{ \mbox{Tr} }

\newcommand{\be}{\begin{equation}}
\newcommand{\ee}{\end{equation}}
\newcommand{\ba}{\begin{array}}
\newcommand{\ea}{\end{array}}
\newcommand{\bea}{\begin{eqnarray}}
\newcommand{\eea}{\end{eqnarray}}

\begin{document}

\title{Number-Phase Wigner Representation for Efficient Stochastic Simulations}

\author{M. R. Hush}
\author{A. R. R. Carvalho}
\affiliation{Department of Quantum Science, Research School of Physics and Engineering, The Australian National University, ACT 0200, Australia}
\author{J. J. Hope}
\affiliation{Australian Centre for Quantum-Atom Optics, Department of Physics, Research School of Physics and Engineering, The Australian National University, ACT 0200, Australia}

\date{\today}


\begin{abstract}
Phase-space representations based on coherent states (P, Q, Wigner) have been successful in the creation of stochastic differential equations (SDEs) for the efficient stochastic simulation of high dimensional quantum systems. However many problems using these techniques remain intractable over long integrations times. We present a number-phase Wigner representation that can be unraveled into SDEs. We demonstrate convergence to the correct solution for an anharmonic oscillator with small dampening for significantly longer than other phase space representations. This process requires an effective sampling of a non-classical probability distribution. We describe and demonstrate a method of achieving this sampling using stochastic weights. 
\end{abstract}


\maketitle

\section{Introduction} \label{sec:intro}

Coherent states have traditionally been used as the basis for the creation of phase-space representations\cite{gardiner:91,milburn:08}.  A major use of these representations has been the production of stochastic differential equations that can greatly reduce the dimensionality of a problem.  For example, a set of $N$ harmonic oscillators which would typically require a density matrix with $D^{2N}$ components (where $D$ is the number of elements you have in your truncated basis) to solve directly can be changed to a set of only $N$ stochastic differential equations. Not only are the memory requirements of the generated stochastic differential equations logarithmically smaller, they also typically integrate much faster than a direct master equation approach \cite{gardiner:01,steel:98,hope:01,norrie:05,johnsson:07,dall:09}.

Unfortunately, some problems have severely limited integration times or cannot be solved at all using standard phase-space methods \cite{gardiner:91,gilchrist:97,sinatra:02}. A coherent state basis is inappropriate for many problems and we demonstrate that there is a class of quantum systems for which a number-phase based alternative will do better. A variety of number-phase `Wigner-like' representations have been created in the past. The primary reason for such a diversity of solutions to a problem, which is quite clear in the coherent state case, is the ambiguity of the phase operator in quantum mechanics. The problem was originally investigated by Dirac when he attempted to canonically quantise the electric field using number and phase as canonical operators, however it was soon found that such an approach leads to irreparable contradictions \cite{dirac:27}.  A more algebraic approach was taken by Susskind and Glowger, which resulted in a phase operator that had many of the required properties but was not Hermitian \cite{susskind:64}. This issue was investigated and improved using a variety of techniques created by Pegg and Barnett~\cite{barnett:86,barnett:89}, and many number-phase-space representations were created using their phase operators~\cite{vaccaro:95,vaccaro:90}. More recently Moya-Cessa has presented an number-phase Wigner representation based on the Susskind and Glogower operator directly \cite{moya-cessa:03} which does not require an extention or truncation of the fock space like Pegg and Barnett's phase operators.  However, all these approaches were primarily motivated with visualising quantum states and lacked the operator correspondences required for the generation of stochastic differential equations. This paper presents a number-phase Wigner representation that can be used to generate stochastic differential equations.  We investigate the properties of this distribution and use it to demonstrate efficient numerical results.

We find that a coherent state in the number-phase Wigner representation has a non-classical probability distribution, where sections of the distribution are negative. Simulation of such states is important for comparative and practical applications of the representation. Thus, we develop a method for sampling non-classical distributions using stochastic weights. Stochastic weights have been used previously to improve convergence of positive P simulations \cite{deuar:01} and in the simulation of stochastic conditional master equations \cite{hush:09}.

\section{Number-Phase Wigner Representation \label{sec:numphwigrep}}

We define our representation in such a way that a master equation can be converted directly to the equation of motion for the phase-space function, similar to Wigner's original formulation \cite{wigner:32}, as opposed to defining it by the ordering of the operators. We take the definition:
\begin{equation}
 W(n,\phi) = \frac{1}{2\pi} \sum_{k=-n}^{n}\langle n + k| \hat{\rho} | n - k \rangle e^{-2i\phi k}, \label{eqn:defnnumwig}
\end{equation}
where $n \in {0,1/2,1,3/2,2...}$, $\phi \in (0,2\pi)$, $\hat{\rho}$ is a density matrix and $| p \rangle$ is a Fock state ($p \in {0,1,2...}$) the eigenstate of the number operator $\hat{n} \ket{p}  = p \ket{p}$.  The sum increases by increments of $1$ (not $1/2$), ensuring that the Fock states are always integer numbers.  While in many previous phase-space representations the domain of the variables is the set of eigenvalues of particular operators, in this case we require $n$ to take half integer values as a convenient way to ensure that the representation is complete. Consider
\begin{eqnarray}
\int d\phi & W(n,\phi) &e^{2im\phi}  \nonumber \\ 
& = & \sum_{k=-n}^n \bra{n+k} \mo{\rho} \ket{n-k} \int d\phi\; e^{-2i(k-m)\phi} \nonumber \\
& = & \bra{n+m} \mo{\rho} \ket{n-m},
\end{eqnarray}
thus $n$ and $m$ must take on half integer values to ensure that that all density matrix elements (such as $\bra{n}\mo{\rho}\ket{n+1}$) can be extracted. 
We can see that this definition has many of the properties expected from a Wigner representation: 
\\
i) $W(n,\phi)$ is always real
\begin{eqnarray}
&W(n,\phi)^*& \nonumber \\
& = & \frac{1}{2\pi} \sum_{k=-n}^{n} \int _0^{2\pi} d\phi \; \langle n - k| \hat{\rho} | n + k \rangle e^{2i\phi k} \nonumber \\
& = & \frac{1}{2\pi} \sum_{k=-n}^{n} \int _0^{2\pi} d\phi \; \langle n + k| \hat{\rho} | n - k \rangle e^{-2i\phi k} \nonumber \\
& = & W(n,\phi);
\end{eqnarray}
ii) Integrating out the phase variable return the expected distribution for number 
\begin{eqnarray}
\int_0^{2\phi} d\phi \; W(n,\phi) = \bra{n} \mo{\rho} \ket{n} = P(n).
\end{eqnarray}
iii) Summing out the number variable returns the expected distribution for phase
\begin{eqnarray}
&\sum_{n=0}^{\infty} W(n,\phi)& \nonumber \\
& = & \frac{1}{2\pi} \sum_{n=0}^{\infty}\sum_{k=-n}^{n} \langle n + k| \hat{\rho} | n - k \rangle e^{-2i\phi k} \nonumber \\
& = & \sum_{a=0}^\infty \sum_{b=0}^\infty \bra{a} \mo{\rho} \ket{b} e^{-i(a-b)\phi} \nonumber \\
& = & \bra{\phi} \mo{\rho} \ket{\phi} = P(\phi),
\end{eqnarray}
where  $a,b \in {0,1,2...}$ and $\ket{\phi}$ is the eigenfunction of  $e^{i\hat{\Phi}} = \sum_{m=0}^{\infty} |n \rangle \langle n+1|$ the exponential phase operator as defined by Susskind and Glogower \cite{susskind:64}. 

The main advantage of this representation is that it is straightforward to derive the operator correspondences between a master equation and a partial differential equation for the evolution of the number-phase representation. For example consider the correspondence for the number operator, replacing $\hat{\rho} \hat{n}$ into our definition (\ref{eqn:defnnumwig})
\begin{eqnarray}
& & \frac{1}{2\pi} \sum_{k=-n}^{n}\langle n + k| \hat{\rho} \hat{n} | n - k \rangle e^{-2i\phi k} \nonumber \\
& = &\frac{1}{2\pi} \sum_{k=-n}^{n}\langle n + k| \hat{\rho} (n-k) | n - k \rangle e^{-2i\phi k} \nonumber \\
& = & (n - \frac{i}{2} \partial_\phi) W(n,\phi) 
\end{eqnarray}
Using similar techniques for the other operators we find:
\begin{eqnarray}
\hat{\rho} \hat{n} & \rightarrow & (n - \frac{i}{2} \partial_\phi)W(n,\phi), \nonumber \\
\hat{n} \hat{\rho} & \rightarrow & (n +\frac{i}{2} \partial_\phi)W(n,\phi), \nonumber \\
\hat{\rho}e^{i\hat{\Phi}} & \rightarrow & e^{i\phi} W(n-\frac{1}{2},\phi), \nonumber \\
(e^{i\hat{\Phi}})^\dag \hat{\rho} & \rightarrow & e^{-i\phi} W(n-\frac{1}{2},\phi), \nonumber \\
\hat{\rho}  (e^{i\hat{\Phi}})^\dag & \rightarrow & e^{-i\phi} \left( W(n+\frac{1}{2},\phi) \right. \nonumber \\
& &\left. - \int_0^{2\pi} d\phi' e^{-2i(\phi - \phi')(n+\frac{1}{2})} W(n+\frac{1}{2},\phi') \right) ,\nonumber \\
e^{i\hat{\Phi}} \hat{\rho} & \rightarrow & e^{i\phi} \left( W(n+\frac{1}{2},\phi) \right. \nonumber \\
& & \left. - \int_0^{2\pi} d\phi' e^{2i(\phi - \phi')(n+\frac{1}{2})} W(n+\frac{1}{2},\phi')\right). 
 \label{eqn:corres}
\end{eqnarray}
The expectation values in the phase operator correspondences can be re-written as correlations with the vacuum as $\int_0^{2\pi} d\phi' e^{-2in \phi'} W(n,\phi') =   \langle 0 |\hat{\rho}| 2n \rangle$. Thus for large fields where occupation of the vacuum is low, we can neglect these terms and our operator correspondences become analogous to the coherent Wigner case where $\hat{x} \rightarrow \hat{n}$ and $\hat{p} \rightarrow \hat{\Phi}$ ($\hat{x}$ and $\hat{p}$ are the position and momentum operators receptively). The canonical commutator and the Hermiticity of the phase operator is restored in the large field limit so this result is not surprising.

As one must sample the distribution before any numerical simulation, it is useful to examine the distribution for some typical physical states. First let's consider a number state $\mo{\rho} = \ket{m}\bra{m}$
\begin{eqnarray}
W_m(n,\phi) & = &  \frac{1}{2\pi} \sum_{k=-n}^{n}  \bra{ n + k } m \rangle \langle m \ket{n - k} e^{-2i\phi k}  \nonumber \\
W_m(n,\phi) & = & \frac{\delta_{n,m}}{2\pi}.
 \end{eqnarray}
 The state is completely defined in number but unknown in phase as one would expect. Next let's consider a coherent state $\hat{\rho} = \ket{\alpha}\bra{\alpha}$
\begin{eqnarray}
W_{\alpha}(n,\phi) = \frac{1}{2\pi} \sum_{k=-n}^n \bra{n+k}  \alpha \rangle \langle \alpha  \ket{n-k} e^{-2i\phi k} \nonumber \\
 =  \frac{1}{2\pi} \sum_{k=-n}^n \sum_{p,q=0}^\infty \frac{\alpha^p{\alpha^*}^q}{\sqrt{p!q!}}\bra{n+k} p \rangle \langle q\ket{n-k} e^{-|\alpha|^2-2i\phi k} \nonumber \\
W_{\alpha}(n,\phi) = \frac{1}{2\pi} \sum_{k=-n}^n \frac{|\alpha|^{2n} e^{-|\alpha|^2+2ik(\arg{\alpha} - \phi)} }{\sqrt{(n+k)!(n-k)!}}.  \label{eqn:costate}
\end{eqnarray}
Finally, we also need to determine the expectation value correspondence. Investigating number and phase operator ordering is probably of little practical use as relevant observables are typically not a simple function of ordered number and phase operators, and re-arranging such functions into a specific ordering is not simple.  Fortunately, we can find a connection between moments of our Wigner function and anti-normally ordered ladder operators. Consider 
\begin{eqnarray}
&  & \int_0^{2\pi} d\phi \sum_{n=0}^\infty \frac{(n+\frac{p+q}{2})! e^{i(q-p)\phi} }{\sqrt{(n+\frac{q-p}{2})!(n+\frac{p-q}{2})!}} W(n,\phi) \nonumber \\
& = & \sum_{n=0}^\infty \sum_{k=-n}^{n} \frac{(n+\frac{p+q}{2})!\bra{n+k}  \mo{\rho} \ket{n - k}  \int_0^{2\pi} d\phi e^{-i\phi(2k+q-p)} }{2\pi \sqrt{(n+\frac{q-p}{2})!(n+\frac{p-q}{2})!}}   \nonumber \\
& = & \sum_{n=\abs{\frac{q-p}{2}}}   \frac{(n+\frac{p+q}{2})!\bra{n+\frac{q-p}{2}}  \mo{\rho} \ket{n - \frac{p-q}{2}}}{\sqrt{(n+\frac{q-p}{2})!(n+\frac{q-p}{2})!}}  \nonumber \\
& = & \sum_{max\{p,q\}}^\infty \frac{n!\bra{n-p} \mo{\rho} \ket{n-q}}{\sqrt{(n-p)!(n-q)!}}  \nonumber \\
& = & \sum_{n=0}^\infty \bra{n} (\mo{a}^\dag)^p \mo{\rho} \mo{a}^q\ket{n} = \Tr[\mo{a}^q(\mo{a}^\dag)^p \mo{\rho} ]
\end{eqnarray}
Note we made a substitution on the 4th line and let $n \rightarrow n + \frac{p-q}{2} $. The identity above is the most straightforward way to make expectation value correspondences for most physically interesting observables.

\section{Non-Classical Probability Sampling} \label{sec:nonclassprobsamp}

In section \ref{sec:numphwigrep} we have found that a coherent state in the number-phase Wigner representation does have some negativity. Only strictly positive probability distributions can be sampled using traditional sampling techniques, so it is necessary to introduce an additional degree of freedom to allow unorthodox averaging. We do this by introducing a weight for each path, and in particular we allow these weights to be negative. Traditionally weights have been used when combining data sets of varying confidence, for example combining the results of different experiments where a different number of measurements have been taken. In this case, weights are kept strictly positive and are a measure of the relative confidence of the data \cite{cochran:37,meier:53}. However, in recent times weights have become more common as a useful tool in a variety of fields: in the simulation of conditional master equations~\cite{hush:09}, and to improve convergence in stochastic simulations of multimode bosonic and fermionic fields~\cite{deuar:02,corney:04}. In some recent applications of weights it has been advantageous to allow weights to take negative values, whether this is because of the weights become negative during the implementation of an algorithm like in the field of engineering~\cite{shi:02,arce:00,fakcharoenphol:06} or for sampling purposes in monte carlo simulation of fermionic systems~\cite{nedjalkov:04,chang:04}.

We present two methods based on weights. The first is a probabilistic method that we will prove is the optimal way of sampling a non-classical probability distribution in general. The second method is a deterministic method, which may have an advantage when the computational requirements of the first method are too high for particular distributions. We also quantify how these two methods can be compared. 

We define weighted averages as follows
\begin{equation}
\overline{f(\mv{x})} = \frac{\sE[\omega f(\mv{x} )]}{\sE[\omega]} \label{eqn:wavgdefn}.
\end{equation}
We begin with the probabilistic method.  Consider a probability $P(\mv{x})$ that is normalised, but not strictly positive. We assume $P(\mv{x})$ can be split into a strictly positive function, $P^+(\mv{x}) \ge 0 \; \forall \; \mv{x}$, and  strictly negative function, $P^-(\mv{x}) \le 0 \; \forall \; x$, as follows
\begin{equation}
P(\mv{x}) = P^+(\mv{x}) + P^-(\mv{x}).
\end{equation}
We further require that the functions $P^{\pm}(\mv{x})$ have finite integral, specifically $C^{\pm} = \int d\mv{x} \;P^{\pm}(\mv{x})$ (note that $C^++C^-=1$).  We can then define two strictly positive normalised probability distributions  as $\tilde{P}^\pm(\mv{x}) = P^{\pm}(\mv{x}) / C^{\pm}$, which can be sampled in a traditional manner.  Now let us consider the average produced if we sample the random variables $x^\pm_i$ from the distributions $\tilde{P}^\pm(\mv{x})$ with rates $\lambda^\pm$, and assign each path a weight $\omega_\pm = C^\pm /\lambda^\pm$  for a total of $N$ samples where $1>\lambda^\pm>0$ and $\lambda^+ + \lambda^- = 1$. For a sufficiently large $N$ we can assume $N\lambda^+$ paths will be sampled from $P^+(\mv{x})$ and $N\lambda^-$ paths will be sampled from $P^-(\mv{x})$, the average produced by this technique, using (\ref{eqn:wavgdefn}), is
\begin{eqnarray}
\overline{f(\mv{x})} & = & \frac{\sum_i^{\lambda^+ N} \omega_+ f(\mv{x}_i^+) + \sum_i^{\lambda^-N} \omega_- f(\mv{x}_i^-)}{\sum_i^{\lambda^+N} \omega^+ + \sum_i^{\lambda^-N} \omega_- } \nonumber \\
& = &  \frac{\sum_i^{\lambda^+N} \frac{C^+}{\lambda^+} \ f(\mv{x}_i^+) + \sum_i^{\lambda^-N} \frac{C^-}{\lambda^-} f(\mv{x}_i^-)}{N}\nonumber \\
\overline{f(\mv{x})}  & = & \sum_i^{\lambda_+ N} \frac{f(\mv{x}^+_i)}{\lambda^+N} C^+ + \sum_i^{\lambda^- N}  \frac{f(\mv{x}_i^-)}{\lambda^- N} C^- \label{eqn:avgfxsum}. 
\end{eqnarray}
In the limit of large $N$ the sum's of random variables in (\ref{eqn:avgfxsum}) converge to the integral of the probability distribution, as required by the definition of a random variable, thus we can replace
\begin{eqnarray}
\overline{f(\mv{x})}  & = & \int d\mv{x} f(\mv{x}) \tilde{P}^+(\mv{x}) C^+ + \int d\mv{x} f(\mv{x}) \tilde{P}^-(\mv{x}) C^- \nonumber \\
& = & \int d\mv{x} f(\mv{x}) ( P^+(\mv{x}) + P^-(\mv{x}) ) \nonumber \\
\overline{f(\mv{x})}  & = & \int d\mv{x} f(\mv{x}) P(\mv{x}). \label{eqn:avgfxint}
\end{eqnarray}
Thus, the non-classical distribution $P(\mv{x})$ is sampled by using the dual distributions and probabilities outlined above, and computing averages using the weighted method. The weights in this method are normalised in the sense that $\sE[\omega]=1$, but this can be scaled arbitrarily if desired.  There is still a large degree of freedom in what we choose for $\lambda^+$ and $C^+$, which needs to be further investigated to determine the optimal sampling technique.  Optimisation requires a measure of the accuracy of the calculated means, which is traditionally measured using the variance.  This approach is validated by the central limit theorem, which can be extended to apply to the weighted mean case.  Assuming the weights themselves can be treated as random variables, and remembering to take into account correlations, we find:
\begin{equation}
\sigma \left(f(\mv{x}) \right)^2 = \frac{\overline{f(\mv{x})}^2}{N} \left(\frac{\sE[\omega^2 f(x)^2]}{\sE[\omega f(x)]^2} - 2 \frac{\sE[\omega^2 f(x)]}{\sE[\omega f(x)]\sE[\omega]} + \frac{\sE[\omega^2]}{\sE[\omega]^2} \right).
\end{equation}
The variance reduces in the limit of large sampling, but to make the most efficient use of resources, we want to minimise the variance for a fixed $N$. As mentioned above, we have a degree of freedom in our choice of both $\lambda^+$ and $C^+$ which can be optimised. The terms that depend on $f(\mv{x})$ will have a convoluted dependence on $P(\mv{x})$ and on the particular observable that is being calculated. The best we can hope achieve is to minimise $\sE[\omega^2]/\sE[\omega]^2$, this will reduce the uncertainty in observables as a whole. Although we acknowledge that for specific observables and $P(\mv{x})$'s this is not necessarily true, from now on we restrict our analysis to minimising $\sE[\omega^2]/\sE[\omega]^2$ with respect to $\lambda^+$ and $C^+$. Taking the derivative of 
\begin{equation}
\frac{\sE[\omega^2]}{\sE[\omega]^2} = \frac{(C^+)^2}{\lambda^+} + \frac{(C^-)^2}{\lambda^-} =  \frac{(C^+)^2}{\lambda^+} + \frac{(1 - C^+)^2}{1- \lambda^+}, \label{eqn:v2mina}
\end{equation}
we find that $\sE[\omega^2]/\sE[\omega]^2$ is minimised when $\lambda^+ = C^+$ or $\lambda^+ = \frac{C^+}{2C^+ - 1} = \frac{C^+}{C^+-C^-}$. The first case  $\lambda^+ = C^+$ is only possible when $C^+=1$ and $C^-=0$ as both $0 \le \lambda^{\pm} \le 1$, this scenario only occurs when the probability distribution $P(x)$ has no negativity and the weighted sampling technique reduces to its traditional counterpart. The second case is of more practical use, remembering $C^+>1$ and $C^- <0$ the expression for  $\lambda^+ = \frac{C^+}{C^+-C^-}$ implies $\lambda^- = \frac{-C^-}{C^+ - C^-}$ which both satisfy $0 \le \lambda^{\pm} \le 1$ thus they can be sampled traditionally. Replacing these into (\ref{eqn:v2mina}) we obtain
\begin{equation}
\frac{\sE[\omega^2]}{\sE[\omega]^2} = (C^+ - C^-)^2
\end{equation}
This shows that the optimal sampling technique is achieved by splitting $P(\mv{x}) = P^+(\mv{x}) + P^-(\mv{x})$ such that $(C^+ - C^-)^2$ is minimised.  This can be achieved by defining $P^{\pm}(x)$ piecewise, bounded by the points where $P(\mv{x})$ crosses the $\mv{x}$ axes.  Then the random variables $x^\pm_i$ are sampled from the distribution $\tilde{P}^\pm(\mv{x})$ with a rate $\lambda^\pm =  \frac{C^\pm}{C^+-C^-}$ and assign each path a weight $\omega_\pm =  (C^+-C^-)$ for a total of $N$ samples.

Of course, if the form of the resulting functions is computationally difficult to sample, a method with sub-optimal sampling might be more efficient overall.  An example of this is a simulation where we wish to sample the coherent state from equation (\ref{eqn:costate}).  Splitting the function into two as outlined above and analytically inverting the parts for sampling is not possible.  While numerically inverting the equations would be an option, this would restrict our random samples to starting on a coarse grid, and the majority of the advantages outlined above would be lost.  Instead we can use a simple deterministic technique to approximate the initial sampling. The additional speed in using this technique allows us to use a finer grid, which will make this technique easier and competitive. 

Consider sampling the function $P(\mv{x})$, assume that the function is zero (or sufficiently close to zero) outside of a volume unit of $V$. Divide this volume into equally spaced grid of $N$ points, then at each point assign the random variable $\mv{x}_i$ the value of the point at that grid and give that path a weight of $\omega_i = V P(\mv{x}_i)$. Using (\ref{eqn:wavgdefn}) we see the average calculated using this technique gives
\begin{equation}
\overline{f(\mv{x})} =  \frac{\sum_i  \Delta_\mv{x} P(\mv{x}_i) f(\mv{x}_i)}{\sum_i \Delta_\mv{x} P(\mv{x}_i)}
\end{equation}
where $\Delta_{\mv{x}} = V/N$. In the limit of large $N$ we can assume that the sums approximate the integrals outlined below
\begin{equation}
\overline{f(\mv{x})} \approx \frac{ \int d\mv{x} P(\mv{x}) f(\mv{x})}{\int  d\mv{x} P(\mv{x}_i)} =  \int d\mv{x} P(\mv{x}) f(\mv{x}),
\end{equation}
as required.  We can investigate the contribution to the variance of a calculated mean 
\begin{equation}
\frac{\sE[\omega^2]}{\sE[\omega]} = \sum_i V \Delta_\mv{x} (P(\mv{x}_i))^2 \approx V \int d\mv{x} (P(\mv{x}))^2 
\end{equation}
Unlike the previous technique we cannot optimise this expression, as $V$ is fixed by the distribution itself. Typically the expression above will be significantly larger than the maximum efficiency achieved using the first technique.  When considering a problem one needs to compare the ratio $\sE[\omega^2]/\sE[\omega]$ for each of the two methods above in order to decide whether the computational difficulty implementing the first method is worth the improvement to the variance. The determinisitic method, if computed as an initial condition before simulating, can also be improved by using the breeding algorithm outlined in paper \cite{hush:09}.

\section{Numerical Simulations}

We now demonstrate the effectiveness of stochastic methods based on our number-phase Wigner representation by simulating a damped anharmonic oscillator. An analytic solution has been calculated for this problem with a coherent state as an initial condition in \cite{milburn:86}, and will be used to compare the accuracy of methods. We will show that in the case of weak damping, the number-phase Wigner representation converges over a significantly longer time interval than both gauge positive-P and truncated Wigner methods. We chose to model an anharmonic oscillator because it is algrebraically analogous to simulating the nonlinear term present in the hamiltonian of a Bose Einstien Condensate. This nonlinear term is typically what limits the convergence of previously developed scalable methods. The addition of dampening was neccesary because the nonlinear term is analytic for the number-phase Wigner representation, thus in the undampened case it would trivially win against other scalable methods. To make the comparison fair we added dampening which is both physically reasonable and non-trivial for the number-phase Wigner representation.

Consider the master equation of an anhamonic oscillator with damping
\begin{equation}
\partial_t \hat{\rho} =  -i \frac{\chi}{2} \left[ (\mo{a}^\dag \mo{a})^2,\rho \right] + \gamma \mathcal{D}[\mo{a}] \hat{\rho}. \label{eqn:mastan}
\end{equation}
Applying the correspondences (\ref{eqn:corres}) we find the number-phase space evolution to be
\begin{eqnarray}
\partial_t W(n,\phi) & = & \partial_\phi \chi n  W(n,\phi)  - \gamma n W(n,\phi) \nonumber \\
& & + \gamma\sqrt{(n+1)^2 + \frac{1}{4} \partial_\phi^2} \, W(n+1,\phi). \label{eqn:num1}
\end{eqnarray}
Unraveling this equation into a set of stochastic differential equations requires the application of two approximations. First, we expand the square root
\begin{equation}
\sqrt{(n+1)^2 + \frac{1}{4} \partial_\phi^2}  =  (n+1) + \frac{\partial_\phi^2}{8(1+n)} + O(\partial_\phi^4), 
\end{equation}
and we ignore terms of order $O(\partial_\phi^4)$ and higher. This approximation improves in the limit of large fields. Under this approximation, equation (\ref{eqn:num1}) becomes
\begin{eqnarray}
\partial_t W(n,\phi) & \approx & \partial_\phi \chi n W(n,\phi) +  \frac{ \gamma \partial_\phi^2}{8(n+1)} W(n+1,\phi) \nonumber \\ 
& &  \gamma ((n+1) W(n+1,\phi) - n W(n,\phi)). \label{eqn:num2}
\end{eqnarray}
We are unable to stochastically unravel this equation due to the second term in (\ref{eqn:num2}), a nonlocal differential, so we make the following addition and subtraction to (\ref{eqn:num2})
\begin{eqnarray}
\partial_t W(n,\phi) & \approx & \partial_\phi \chi n W(n,\phi) +  \frac{\gamma \partial_\phi^2}{8(n+1)} W(n,\phi) \nonumber \\ 
& &  \gamma((n+1) W(n+1,\phi) - n W(n,\phi)) \nonumber \\
& &  \frac{\gamma\partial_\phi^2}{8(n+1)} (W(n+1,\phi) - W(n,\phi)).  \label{eqn:num3}
\end{eqnarray}
For a sufficiently smooth function, we can assume the term on the last line of ($\ref{eqn:num3}$) will be small. For the coherent state, which we intend to use as an initial condition for our problem, this is quite a reasonable approximation. Ignoring this term, we make a second approximation reducing (\ref{eqn:num3}) to
\begin{eqnarray}
\partial_t W(n,\phi) & \approx &  \partial_\phi \chi n W(n,\phi) +   \frac{\gamma\partial_\phi^2}{8(n+1)} W(n,\phi) \nonumber \\  
& & + \gamma((n+1) W(n+1,\phi)- n W(n,\phi)). \label{eqn:numfin}
\end{eqnarray}
We can now unravel this equation into a set of stochastic stochastic differential equations.  With some minor inspection one can recognise that the first term and second terms correspond to drift and diffusion in $\phi$ respectively, and the last two terms correspond to a Poissonian jump in $n$.  
\begin{eqnarray}
d\phi_t & = & -\chi m \; dt + \frac{\sqrt{\gamma}}{2\sqrt{n_t+1}} \; dW_t, \nonumber \\
dn_t & = & -dN_t. \label{eqn:cnumphan}
\end{eqnarray}
Here $dW_t$ is an Ito Wiener process and $dN_t$ is a Poisson jump process with a rate $\sE[dN_t] = \gamma \sE[n] dt$. We integrate (\ref{eqn:cnumphan}) then compare the expectation value of position to two other leading scaleable stochastic differential methods: Gauge-$P^+$ \cite{deuar:01,plimak:03} and truncated Wigner \cite{norrie:05}. Gauge-$P^+$ is the longest-converging exact stochastic method for an anharmonic oscillator, for which it can generate the following equations of motion:
\begin{eqnarray}
d\phi_t & = & \left(\chi(n(1-i) - 2G_1(t+i))-\frac{\gamma}{2} (1+i)\right) dt + \sqrt{2\chi} dW_t^1, \nonumber \\
d\psi_t & = & \left(\chi(n^*(1-i) - 2G_2(T-i))-\frac{\gamma}{2} (1+i)\right) dt + \sqrt{2\chi} dW_t^2, \nonumber \\
d\tilde{\theta}_t & = & -2T(G_1^2 + G_2^2)dt + \sqrt{2} (G_2dW_t^2 - G_1dW_t^1).
\end{eqnarray}
where $\alpha = e^{(\frac{1-i}{2})\phi_t}$ and $\beta = e^{(\frac{1-i}{2}) \psi_t}$ are the stochastic coherent amplitude and complex conjugate.  These are complex conjugate at the beginning but may diverge afterwards.  We also have $n = \alpha\beta^*$, $G_1 = 0.005[i(n^*-n)/2 - (n^* + n)/2 + \abs{\alpha}^2]$, $G_2 = 0.005[i(n^*-n)/2 + (n+n^*)/2 - \abs{\beta}^2]$, and $T = \tan(\tilde{\theta}_t)$. Position is calculated using $\overline{x}= \frac{1}{\sqrt{2}} \sE[\Re[\omega(\beta^* + \alpha)]]/\sE[\omega]$ where $\omega = 2 e^{\Re[\alpha \beta^*]} \cos (\tilde{\theta}_t)$. Other correspondences and more details are presented in \cite{deuar:01}. The next stochastic differential equation technique we consider is the coherent truncated Wigner method. It is an approximate method that assumes higher order derivates beyond diffusion can be safely ignored.  It has a typically longer numerical convergence time than Gauge $P^+$ however, due to the uncontrolled approximation, there is no indication when the numerical solution begins to diverge from the true solution. The stochastic differential equations of motion are:
 \begin{equation}
 d\alpha = \left(i\chi\alpha\left(\frac{1}{2} - \abs{\alpha}^2\right) - \frac{\gamma}{2} \alpha\right) dt + \frac{\sqrt{\gamma}}{2} (dW_t^1 + i dW_t^2).
 \end{equation}
The observable is given by $\overline{x} = \frac{1}{\sqrt{2}} \sE[\alpha + \alpha^*]$.  
\begin{figure}[htb]
\includegraphics[scale=0.61]{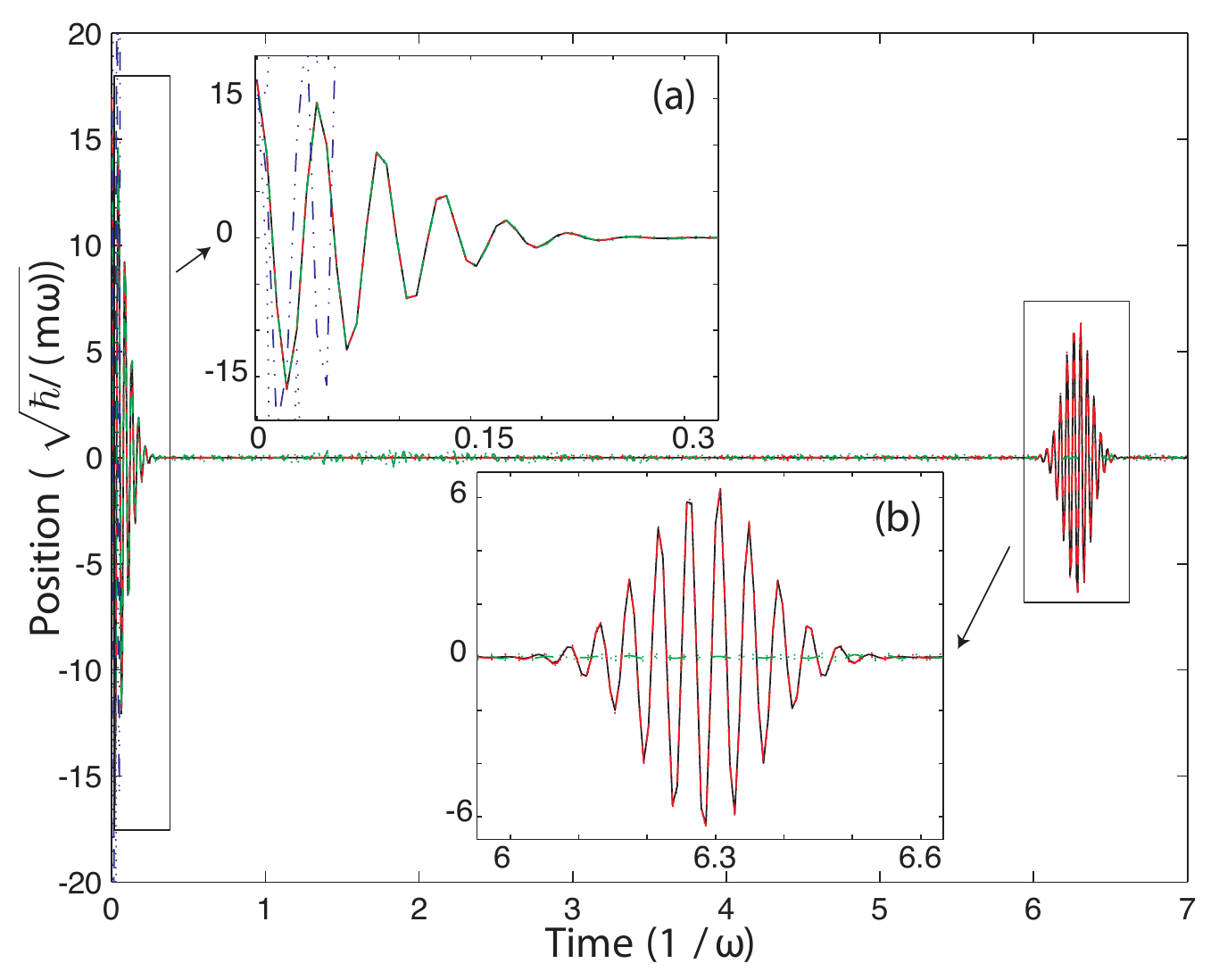}
\caption{(Color online) Position vs. time for a dampened anharmonic oscillator ($\chi = 1$ and $\gamma = 0.001$) integrated using the number-phase Wigner representation in red (dashed), truncated Wigner representation in green (dot dashed), gauge $P^+$ in blue (dot dot dashed) and for comparison the analytic solution in black (solid).  Sampling uncertainties are plotted with dotted lines in the respective colour of the numerical solution. Inset (a) shows the short time period gauge $P^+$ matches the analytic solution until divergent infinite paths dominate the evolution. The coherent Wigner representation converges for the first section of de-phasing in inset (a), however inset (b) shows that it does not converge during the dampened ressurection of the coherent state. The method based on the number-phase Wigner representation converges over the entire integration period. The numerical integration was performed by using the open source software package xpdeint, which is a new version of the xmds package \cite{xmds:01}}
\label{fig:comp}
\end{figure}

The simulations using the different integration methods are presented in Fig.~\ref{fig:comp}. As we can see, the number-phase Wigner representation converges for significantly longer than the other stochastic differential equations and reconstructs the dampened revival of the coherent states successfully. This suggests the number-phase Wigner representation has the possibility to integrate previously difficult quantum problems over significantly longer integration times than other coherent based stochastic differential equations. The anharmonic oscillator example is of particular importance to the field of Bose-Einstein condensation as the anharmonic term has analogous algebraic properties to the dominant nonlinear term present in the many-body master equation.

\section{Conclusion}
We presented a number-phase Wigner representation that is suitable for the investigation of quantum dynamics in phase-space. Direct operator correspondences were derived, allowing a master equation to be written as partial differential equations for the number-phase quasi-probability distribution. Under some approximations, these equations can be unravelled into a set of low-dimensional stochastic equations, hence reducing the dimensionality of the system. This stochastic method was used to model the damped anharmonic oscillator, and it was shown that this method converged dramatically longer than the truncated Wigner and Gauge-$P^+$ methods. This result therefore suggests that a stochastic method based on our number-phase Wigner representation  may be of use for currently difficult problems where number-conserving nonlinear terms dominate the dynamics.

\section{Acknowledgements}
This research was supported under Australian Research Council's Discovery Projects funding scheme (project number DP0556073) and the Australian Research Council Centre of Excellence for Quantum-Atom Optics (ACQAO). We acknowledge the use of CPU time at the National Computational Infrastructure National Facility and thank Graham Dennis for his help with simulations.

\bibliography{references}

\end{document}